\documentclass[12pt]{article}
\usepackage{graphicx,amssymb,amsmath}

\newcommand\pubnumber{DPF2015-146}
\newcommand\pubdate{}

\def\Title#1{\begin{center} {\Large #1 } \end{center}}
\def\Author#1{\begin{center}{ \sc #1} \end{center}}
\def\Address#1{\begin{center}{ \it #1} \end{center}}

\newcommand\pubblock{\rightline{\begin{tabular}{l} \pubnumber\\
         \pubdate  \end{tabular}}}
\newenvironment{Abstract}{\begin{quotation}  }{\end{quotation}}
\newenvironment{Presented}{\begin{quotation} \begin{center} 
             PRESENTED AT\end{center}\bigskip 
      \begin{center}\begin{large}}{\end{large}\end{center} \end{quotation}}
\def\Acknowledgments{\bigskip  \bigskip \begin{center} \begin{large}
             \bf ACKNOWLEDGMENTS \end{large}\end{center}}

\def\beq{\begin{equation}}
\def\eeq#1{\label{#1}\end{equation}}
\def\eeqn{\end{equation}}

\def\leqn#1{(\ref{#1})}

\def\etal{{\it et al.}}

\begin{document}
\begin{titlepage}
\pubblock

\vfill
\Title{Bound on the variation in the fine structure constant implied by Oklo data}
\vfill
\Author{Leila Hamdan\footnote{Presenter} and Edward D. Davis\footnote{Corresponding author. Email: edward.davis@ku.edu.kw}}
\Address{Department of Physics\\ Kuwait University, P.O. Box 5969, Safat 13060, KUWAIT}
\vfill
\begin{Abstract}
Dynamical models of dark energy can imply that the fine structure constant $\alpha$ varies over cosmological time scales. Data 
on shifts in resonance energies $E_r$ from the Oklo natural fission reactor have been used to place restrictive bounds on the 
change in $\alpha$ over the last 1.8 billion years.  We review the uncertainties in these analyses, focussing on corrections to the 
standard estimate of $k_\alpha\!=\!\alpha\,dE_r/d\alpha$ due to Damour and Dyson. Guided, in part, by the best practice for 
assessing systematic errors in theoretical estimates spelt out by Dobaczewski \etal~[in J. Phys. G: Nucl. Part. Phys. {\bf 41}, 
074001 (2014)], we compute these corrections in a variety of models tuned to reproduce existing nuclear data. Although the net 
correction is uncertain to within a factor of 2 or 3, it constitutes at most no more than 25\% of the Damour-Dyson estimate of 
$k_\alpha$. Making similar allowances for the uncertainties in the modeling of the operation of the Oklo reactors, we conclude 
that the relative change in $\alpha$ since the Oklo reactors were last active (redshift $z\simeq 0.14$) is less than $\sim 10$ 
parts per billion. To illustrate the utility of this bound at low-$z$, we consider its implications for the string theory-inspired runaway 
dilaton model of Damour, Piazza and Veneziano.
\end{Abstract}
\vfill
\begin{Presented}
DPF 2015\\
The Meeting of the American Physical Society\\
Division of Particles and Fields\\
Ann Arbor, Michigan, August 4--8, 2015\\
\end{Presented}
\vfill
\end{titlepage}
\def\thefootnote{\fnsymbol{footnote}}
\setcounter{footnote}{0}

\section{Introduction}

In the August, 2015 issue of Physics Today,  Frank Wilczek, reviewing  a new anthology of 
Freeman Dyson's papers, writes that 
``Dyson's paper, with Thibault Damour, placing empirical limits on the possible time variation of the fine-structure and other 
fundamental `constants' is a gem within \emph{Birds and Frogs}."
Since its appearance, Damour and Dyson's paper~\cite{NuclPhysB.480.37} has provided \emph{the\/} basis for relating 
Oklo data to the fine-structure constant $\alpha$, but, in a recent paper~\cite{PhysRevC.92.014319}, we have reconsidered the 
foundations of their method. This contribution to DPF 2015 is a digest of some of the principal ideas and results of our paper,
but it is a more visual presentation, with diagrams generated specifically for DPF 2015  (Ref.~\cite{PhysRevC.92.014319} does 
not contain any figures!). In addition, we briefly take up a topic not discussed in Ref.~\cite{PhysRevC.92.014319}, namely the 
implications of our findings for recent predictions~\cite{PhysLettB.743.377} of the redshift dependence of $\alpha$ in
the runaway dilaton model~\cite{PhysRevD.66.046007}.
 
\section{What is Oklo and why is it of interest?}

There is an extensive body of work on the Oklo phenomenon, the inactive natural nuclear fission reactors discovered in 1972
by the CEA (see Ref.~\cite{IntJModPhysE.23.1430007} for a recent review of the literature). Of most interest to us
is the fact that these reactors were last active about 1.8 to 2 billion years ago. Thus, Oklo geochemical data provide us with
a record of nuclear processes, specifically neutron capture by complex nuclei, at redshift $z\simeq 0.14$.

As first recognised by Alexander I. Shlyakther~\cite{Nature.264.340}, there would have been Oklo capture cross-sections 
which were dominated by the properties of a single compound-nucleus resonance and any difference in such a 
cross-section from its present-day value can be translated into a change $\Delta_r$ in the resonance energy $E_r$ over the
intervening time. Shlyakhter indicated through simple back-of-envelope estimates how a bound on $\Delta_r$
could be used to constrain the change in the dimensionless strengths of the strong, weak and electromagnetic interactions 
over the same time interval.

In work to improve on Shlyakther's estimates, the reaction $\text{n}+{}^{149}\text{Sm}$ has received most attention, primarily
because the cross section for thermal neutron capture is huge (with a resonance just above threshold at $97.3\,\text{meV}$),
offering the best possibility for the determination of $\Delta_r$ (a small change in $E_r$ manifests itself in a 
dramatic change in the capture rate). The information on $\Delta_r$ extracted also depends on the modelling of the neutron 
spectrum within the Oklo reactors. Beginning with Ref.~\cite{NuclPhysB.480.37}, there have been several independent studies
of this issue~\cite{Oklo,PhysRevC.74.024607}
and more work 
would be appreciated! Nevertheless, despite the many uncertainties, the 3 tightest bounds on $\Delta_r$ inferred from Sm data 
all agree to within a factor of 2 with the result of Ref.~\cite{PhysRevC.74.024607}, which we, henceforth, adopt: $\Delta_r =
7.2\pm9.4\,\text{meV}$.

\section{How to extract a bound on {\protect$\Delta\alpha\equiv\alpha_\text{Oklo}-\alpha_\text{now}$}}

The Damour-Dyson method is based on two inequalities. First, there is the plausible assumption that the \emph{net} change in 
$\Delta_r$ exceeds the change due to $\Delta\alpha$ alone, or\footnote{A more complete 
treatment~\cite{PhysRevC.92.014319} of the relation between $\Delta_r$ and $\Delta\alpha$ includes
the effect of a change in the ratio $X_q$ of the average light quark mass $m_q=\tfrac{1}{2}(m_u+m_d)$ to the QCD mass scale 
$\Lambda$; Eq.~\leqn1 is replaced by 
\[
               |\Delta_r| = \left| k_\alpha \frac{\Delta \alpha}{\alpha_\text{now}}  + 
                                            k_q \frac{\Delta X}{X_\text{now}}   \right| 
                               \ge \left| \gamma \nu - 1\right|\, |k_\alpha |\, \frac{|\Delta \alpha |}{\alpha_\text{now}}     ,
\]
where $ \gamma\equiv\left| k_q/k_\alpha \right|$ and  $\nu\equiv\left| \left. \Delta X /X_\text{now}\right/  \Delta\alpha/
\alpha_\text{now} \right| $: as estimates~\cite{PhysRevC.92.014319,FewBodySyst.56.431} of $k_q$ and $k_\alpha$ imply that
$|\gamma|\gtrsim 4$, a sufficient condition for the validity of Eq.~\leqn1 is that the BSM input $\nu>\tfrac{1}{2}$. This 
would appear to be a weak restriction, respected by all existing phenomenological determinations of $\nu$.}
\beq
              |\Delta_r| \ge |k_\alpha|\frac{|\Delta \alpha|}{\alpha_\text{now}} ,
\eeq1
where $k_\alpha \equiv \left. dE_r\right/d\ln\alpha$. It follows that a lower bound to $|k_\alpha|$ is enough to set 
an upper bound on $|\Delta\alpha|$. Via the Hellmann-Feynman theorem, $k_\alpha$ can 
be related to a difference in nuclear coulomb energies~\cite{IntJModPhysE.23.1430007}, and,  by discarding the 
small exchange contribution to nuclear coulomb energies and adroit use of Green's second identity, a second inequality can
be derived~\cite{NuclPhysB.480.37}, namely, the upper bound 
\beq
              k_\alpha \le \int V_\star (\rho_{150^\star} - \rho_{149}) d^3 r,
\eeq2
where $V_\star$ is the electrostatic potential of the excited compound nucleus ${}^{150}$Sm, $\rho_{150^\star}$ is its charge 
density, and $\rho_{149}$ is the ground state charge density of ${}^{149}$Sm. The righthand side of Eq.~\leqn2 proves to be
negative, and so its magnitude is a candidate for the desired lower bound to $|k_\alpha|$.
 
In the evaluation of the righthand side of Eq.~\leqn2, Damour and Dyson make two uncontrolled approximations. They take
the charge distribution of the ${}^{150}$Sm compound nucleus to be a sphere of uniform density, and set
\beq
   V_\star = \frac{Z\, e}{2 R^3} (3 R^2 - r^2)
\eeq4
for \emph{all\/} radial distances $r$ from the center of the sphere ($Z\,e$ is the charge within the sphere and 
$R$ is its radius), meaning 
that, for any choices of charge densities $\rho_1$ and $\rho_2$ (normalised to $Z\, e$),
\[
  \int V_\star ( \rho_1 - \rho_2 ) d^3 r = - Z\,e\; \big(  \langle r^2 \rangle_1 -  \langle r^2 \rangle_2 \big) ,
\]
where  $\langle r^2\rangle_i$ denotes the mean square radial moment of the charge density $\rho_i$. They
further replace the mean square radial moment for the compound nucleus by its value for the ground state of 
${}^{150}$Sm ($\langle r^2\rangle_{150^\star}  \rightarrow\langle r^2\rangle_{150} $) to obtain the final result
\beq
 k_\alpha \le k^{DD} \equiv - \frac{(Z\, e)^2}{2R^3} \big( \langle r^2\rangle_{150} - \langle r^2\rangle_{149} \big) .
\eeq3
The seductive appeal of the Damour-Dyson estimate $k^{DD}$ of a bound on $k_\alpha$ is that it can be calculated
by reference to experimental data (on isotope shifts and equivalent rms charge radii) alone. It is unfortunate then that, in 
computing $k^{DD}$, Damour and Dyson choose an unphysically large value of $R$ ($8.11\,\text{fm}$ instead of $6.50\pm 0.01\,
\text{fm}$ from tabulations of charge radii). Our preferred value of $k^{DD}$ is $-2.51\pm 0.20\,\text{MeV}$ as 
opposed to the significantly smaller value of $-1.1\pm 0.1\,\text{MeV}$ advocated by Damour and Dyson.

{
\begin{figure}[b!]\centering
\includegraphics[width=6in]{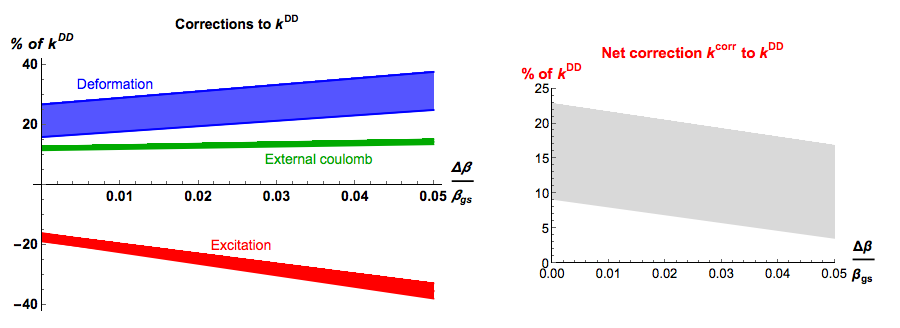}
\caption{\small 
Individual corrections to $k^{DD}$ and the \emph{net\/} correction $k^{corr}$. The bands plotted delineate the range of 
values obtained with the four different models of densities considered.}
\label{fg:Results}
\end{figure}
}

We can identify three nuclear physics corrections to the result in Eq.~\leqn3: an excitation correction, an external coulomb
correction, and a deformation correction. The excitation and external coulomb corrections compensate, respectively, for the 
replacement of $\langle r^2\rangle_{150^\star}$  by $\langle r^2\rangle_{150} $ and for the use throughout all space of Eq.~\leqn4,
a result appropriate to the \emph{inside\/} of a sphere of uniform volume charge density.
The deformation correction accommodates the fact that both the ground state of ${}^{149}$Sm and the compound nucleus 
state of ${}^{150}$Sm excited have prolate deformations. 

We need more realistic charge densities to compute these corrections. We have
employed deformed Fermi profiles fitted to the output of Hartree-Fock+BCS calculations and developed a method for 
including the increase in surface diffuseness with excitation energy. Our results for four different models are plotted 
in Fig.~\ref{fg:Results} for a reasonable range of $\Delta\beta=\beta_\star - \beta_\text{gs}$, where 
$\beta_\star$ and $\beta_\text{gs}$ are the quadrupole deformation parameters for the compound nucleus state excited in 
${}^{150}$Sm and the ground state of ${}^{150}$Sm, respectively. There are significant cancellations between the excitation 
and deformation corrections with the \emph{net\/} correction $k^\text{corr}$ never exceeding 25\% of $k^{DD}$.

\begin{samepage}

On the basis~\cite{JPhysG-NuclPartPhys.41.074001} of the mean and the scatter of estimates of 
$k^\text{corr}$, we conclude that $k^\text{corr} = 0.33\pm 0.16\,\text{MeV}$ or, combining the errors of $k^{DD}$ 
and $k^\text{corr}$ in quadrature, 
\[
 k_\alpha \le k_\text{Bd} \equiv k^{DD} + k^\text{corr} = -2.18 \pm 0.26\,\text{MeV} .
\] 
Despite an uncertainty in $k^\text{corr}$ of about 50\%, the error in the bound on $k_\alpha$ is only a little over 10\% or so.

From Eqs.~\leqn1 and \leqn2, $|\Delta\alpha|/\alpha_\text{now}$ is bounded by $|\Delta_r|/(-k_\text{Bd})$. If we 
interpret $\zeta = -\Delta_r/k_\text{Bd}$ as the ratio of two gaussian random variables of known mean and standard deviation,
then it turns out that $\zeta$ itself is, to an excellent approximation, also a gaussian random variable. Its probability density 
is shown in Fig.~\ref{fg:Gaussian}, along with, for the sake of illustration, the area (in pink) associated with the 68\% 
C.L. upper bound on $|\zeta|$. Using the distribution of $\zeta$, we determine the bound of
\beq
     \frac{|\Delta\alpha|}{\alpha_\text{now}} < 1.1\times 10^{-8}\hspace*{1cm} (95\%\ \text{C.L.}).
\eeq5

{
\begin{figure}[b!]\centering
\begin{minipage}[c]{0.53\textwidth}\centering\vspace{0pt}
\includegraphics[width=\textwidth]{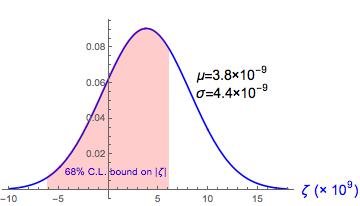}
\caption{\small 
Gaussian distribution of $\displaystyle\zeta\equiv-\frac{\Delta_r}{k_\text{Bd}}$.}
\label{fg:Gaussian}\end{minipage}\hspace*{\fill}
\begin{minipage}[c]{0.43\textwidth}\centering\vspace{0pt}
\includegraphics[width=\textwidth]{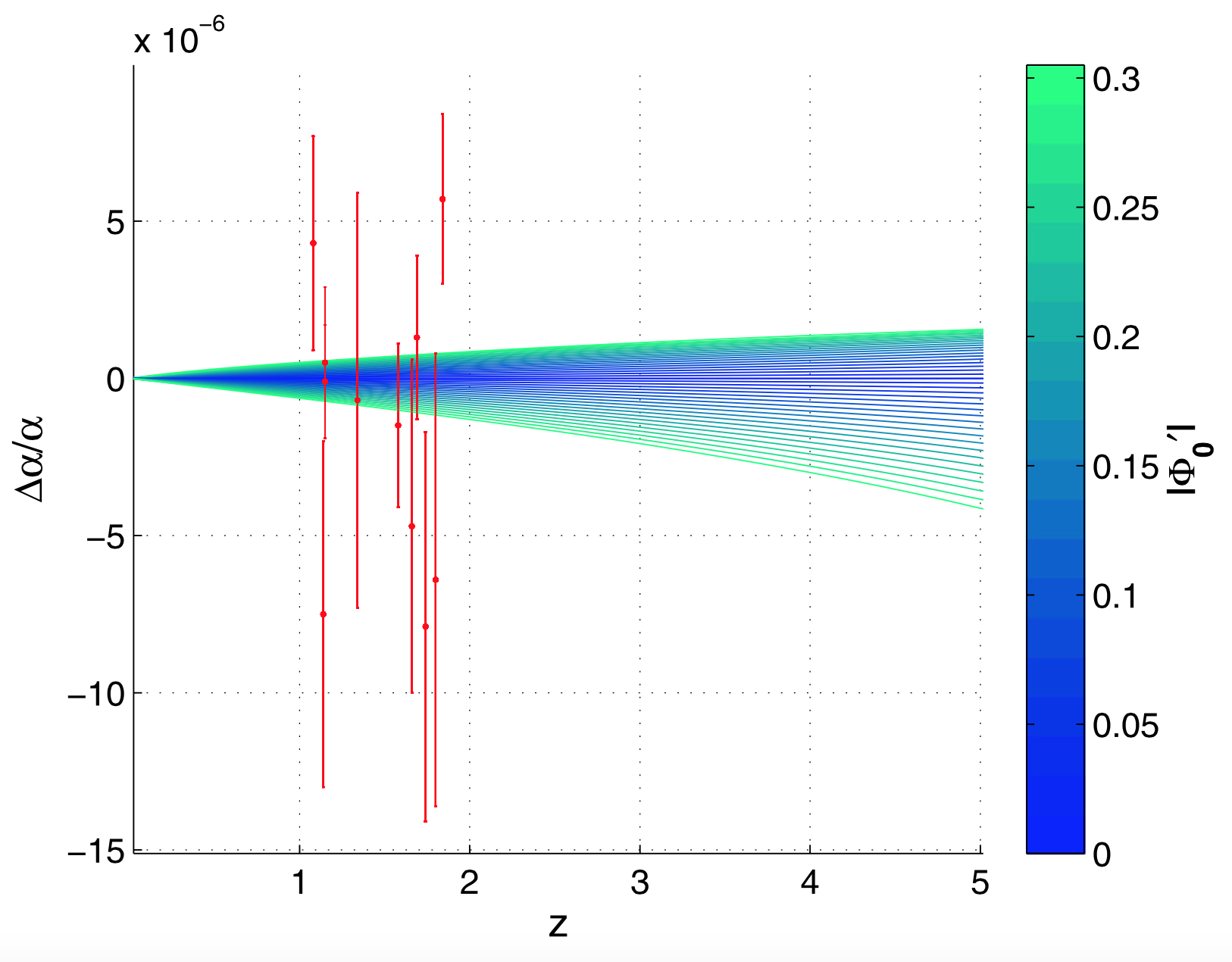}
\caption{\small 
$\Delta\alpha(z)$ in the model of Ref.~\cite{PhysRevD.66.046007} (taken from Fig.~2 of
Ref.~\cite{PhysLettB.743.377}).}
\label{fg:AlphaEvol}
\end{minipage}
\end{figure}
}

\section{Implications for dark cosmology}

Figure~\ref{fg:AlphaEvol} depicts the relation between the redshift dependence of $\alpha$ and the current ``speed'' 
$\Phi_0^\prime$ of the dilaton field in the model of Ref.~\cite{PhysRevD.66.046007} for the updated parameter values of 
Ref.~\cite{PhysLettB.743.377}. Setting $z=0.14$ and $|\alpha_\text{had}|=10^{-4}$ (from Ref.~\cite{PhysLettB.743.377}) in 
\[
  \frac{|\Delta\alpha|}{\alpha}\simeq  \frac{|\alpha_\text{had}|}{40} |\Phi_0^\prime| \ln(1+z) ,
\]
where $\alpha_\text{had}$ denotes the coupling of the dilaton to hadronic matter, our bound on $|\Delta \alpha|$ in Eq.~\leqn5
implies that 
$|\Phi_0^\prime|\lesssim 0.03$. 
Inspection of Fig.~\ref{fg:AlphaEvol} suggests that, for this range of dilation speeds
and the redshifts shown,
the difference in the $z$-dependence of $\alpha$ for $\Lambda$CDM and for the dilaton model will be undetectable.

\end{samepage}

\section{Conclusion}

There are 3 principal conclusions: i) despite suggestions to the opposite in the literature, the (revised) Damour-Dyson estimate of 
$k_\alpha$ is reliable for order of magnitude purposes; ii) at 95\% C.L., 
$|\alpha_\text{Oklo}-\alpha_\text{now}|/\alpha_\text{now} < 1.1\times 10^{-8}$, and; 
(iii) our bound on $|\alpha_\text{Oklo}-\alpha_\text{now}|$ implies that, for redshifts $z<5$, the behaviour of $\alpha$ in the
runaway dilaton model and in $\Lambda$CDM is almost indistinguishable.

\Acknowledgments {
E.D.D. thanks the  Institute for Nuclear Theory at the University of Washington for its hospitality and the Department of Energy
for partial support during preparation of this manuscript. He also thanks INT-15-3 participants Li-Sheng Geng, 
Felipe J. Llanes-Estrada and, especially, John Behr for their interest, and Christopher Gould for bringing F. Wilczek's review of
\emph{Birds and Frogs\/} to his attention.}

\end{document}